%Paper: hep-th/9403002
%From: farina@vms1.nce.ufrj.br
%Date: Tue, 1 Mar 1994 13:42:41 -0300
%Date (revised): Sat, 26 Mar 1994 08:43:23 -0300

\magnification=1200
\hsize=15.5 true cm
\vsize=22 true cm
\baselineskip=18pt
\centerline{\bf SCHWINGER'S FORMULA AND THE PARTITION FUNCTION}
\bigskip
\centerline{ \bf FOR THE BOSONIC AND FERMIONIC HARMONIC OSCILLATOR}
\vskip 2.0cm
\centerline {\it L. C. Albuquerque, C. Farina\footnote*{e-mail: farina@vms1.
nce.ufrj.br} and S. J. Rabello}
\vskip 1.0cm
\centerline{Instituto de F\'\i sica}
\centerline{Universidade Federal do Rio de Janeiro}
\centerline{Rio de Janeiro, R.J.  CEP - 21.945-970 - Brazil}
\vskip 4.0cm
\centerline{\bf Abstract}
\bigskip

We use Schwinger's formula, introduced by himself in the early fifties to
compute effective actions for QED, and recently applied to the Casimir
effect, to obtain the partition functions for both the bosonic and fermionic
harmonic oscillator.
\vfill\eject

The computation of the partition function of the usual harmonic oscillator is
probably one of the most elementary exercises in a Statistical Mechanics
course. There are many ways of making this calculation, and undoubtedly the
easiest is the direct one, that is,
$$\eqalignno{Z(\beta)&:=Tr\, e^{-\beta \hat{H}}\cr
&=\sum_{n=0}^\infty\, e^{-\beta(n+{1\over2})\omega}\cr
&={e^{-{\beta\omega\over2}}\over 1-e^{-\beta\omega}}\cr
&={1\over 2\sinh ({\beta\omega\over2})},&(1)\cr}$$
where we simply used that the eigenvalues of the Hamiltonian operator for the
harmonic oscillator are given by $(n+{1\over2})\omega$, with $n=0,1,2,...$
(we are using $\hbar=1$) and summed the infinite terms of a geometric series.

However, it is exactly the simplicity of handling with this example that makes
it a perfect \lq\lq laboratory" to test or develop other methods of
computation,
as for instance, the path integral method$^{(1)}$, the Green function
method$^{(2)}$, etc..

Our purpose in this letter is to apply a formula invented by Schwinger in
1951$^{(3)}$ to compute effective actions for QED, and recently applied with
success in the computation of the Casimir energy for both the massless and
massive scalar field$^{(4-6)}$, to obtain not only the result of equation (1),
but also the partition function for a fermionic harmonic oscillator. Curious
as it may seem, this approach has never appeared in the literature.

Let us start with the bosonic case. It is well known that the corresponding
partition function can be written as$^{(7)}$
$$Z(\beta)=\det\;^{-{1\over2}}\, (\omega^2-\partial_\tau^2)\vert_{{\cal F}_p}
,\eqno(2)$$
where the subscript ${\cal F}_p$ means that the operator $\omega^2-
\partial_\tau^2$ acts only on a set of functions which are periodic,
with period $\beta$. In ref. {\bf [7]}, Gibbons used the generalized
$\zeta$-function method to compute such a determinant. Here, we shall use
Schwinger's formula (deduced in the {\bf Appendix A}):
$$\ln Z(\beta)={1\over2}Tr\int_0^\infty\, ds\, s^{\nu-1}\,
e^{-is{\hat L}_\omega},\eqno(3)$$
where ${\hat L}_\omega:=\omega^2-\partial_\tau^2$ and we chose the
regularization based on the analytical continuation, instead of Schwinger's
one.

For periodic boundary conditions we have
$$Tr\, e^{-is{\hat L}_\omega}\; =\sum_{n=-\infty}^\infty\,
e^{-is[\omega^2+n^2({2\pi\over\beta})^2]},\eqno(4)$$
so that
$$\eqalignno{\ln Z(\beta)&={1\over2}\sum_{n=-\infty}^\infty\int_0^\infty
\, ds\, s^{\nu-1}\, e^{-is[\omega^2+(n{2\pi\over\beta})^2]}\cr
&={1\over2}\Gamma(\nu)\sum_{n=-\infty}^\infty\left[ \omega^2+
{\left( n{2\pi\over\beta}\right)}^2\right]^{-\nu}\cr
&={1\over2} \omega^{-2\nu}\Gamma(\nu)+
\left({\beta\over 2\pi}\right)^{2\nu}\Gamma(\nu)E_1^{\mu^2}(\nu,1)
,&(5)\cr}$$
where we introduced the one-dimensional inhomogeneous Epstein function
$$E_1^{\mu^2}(\nu,1):=\sum_{n=1}^\infty\left( n^2+\mu^2\right)^{-\nu}
\;\;\; ;\;\;\;Re\,\nu>{1\over2},\eqno(6)$$
defined $\mu={\beta\omega\over 2\pi}$ and used the well known integral
representation of the Euler Gamma function $\int_0^\infty
\, dt\, t^{\nu-1}\, e^{-\alpha t}=\alpha^{-\nu}\Gamma(\nu)$.

Although the above series converges only for $Re\,\nu>{1\over2}$,
it can be analytically continued to a meromorphic function in the whole
complex plane given by$^{(8)}$
$$\eqalignno{E_1^{\mu^2}(\nu,1)&=-{1\over 2\mu^{2\nu}}+{\sqrt{\pi}
\Gamma(\nu-{1\over2})\over 2\Gamma(\nu)\mu^{2\nu-1}}\cr
&+{2\sqrt{\pi}\over \Gamma(\nu)}\sum_{n=1}^\infty\left({n\pi\over\mu}
\right)^{\nu-{1\over2}}\, K_{\nu-{1\over2}}\,(2\pi n\mu).&(7)\cr}$$

It is worth noting that the structure of poles of $E_1^{\mu^2}(\nu,1)$ is
governed by the poles of $\Gamma(\nu-{1\over2})$. Hence, they are located at
$\nu={1\over2},\;-{1\over2},\;-{3\over2},...,$ and so on. As we see,
$E_1^{\mu^2}(\nu,1)$ is analytic at the origin.

Substituting (7) into (5), taking the limit $\nu\rightarrow 0$ and observing
that the divergent terms cancel without any further subtraction, we get
$$\ln Z(\beta)=-{\omega\beta\over2}
+2\sqrt{{\omega\beta\over 2\pi}}\sum_{n=1}^\infty
\sqrt{{1\over n}}\, K_{-{1\over2}}\,(n\omega\beta).\eqno(8)$$

In order to compute the summation on the r.h.s. of (8), we appeal to the
formula$^{(9)}$
$$K_{-{1\over2}}\,(n\omega\beta)=\sqrt{{\pi\over 2n\omega\beta}}\,
e^{-n\omega\beta}.\eqno(9)$$
Inserting (9) into (8), and using equation (B.1) (see {\bf Appendix B}),
we obtain the final result
$$\ln Z(\beta)=-\ln\left[ 2\sinh\left({\omega\beta\over2}\right)
\right]\;\;\;\;\Longrightarrow\;\;\; Z(\beta)={1\over 2\sinh\left(
{\omega\beta\over2}\right)},\eqno(10)$$
in perfect agreement with (1).

For the fermionic case, it can be shown that
$$Z^f(\beta)=\det\;^{+1}(\omega^2-\partial_\tau^2)\vert_{{\cal
F}_a},\eqno(11)$$
where the subscript ${{\cal F}_a}$ now means that the operator
${\hat L}_\omega$ acts only on a set of antiperiodic functions. As a
consequence,
the eigenvalues turn to be $\lambda_n=\omega^2+\left( p{\pi\over\beta}
\right)^2$, with $p$ an odd integer. Then, using these eigenvalues in the
analogue of equation (3), but remembering that for the fermionic case, instead
of the factor ${1\over2}$ we must write $(-1)$, we get
$$\ln Z^f(\beta)=-\Gamma(\nu)\sum_{p=odd}\, \left[\omega^2+\left( p{\pi
\over\beta}\right)^2\right]^{-\nu}.\eqno(12)$$
Adding and subtracting $\Gamma(\nu)\sum_{p=even}\,\left[\omega^2
+\left(p{2\pi\over\beta}\right)^2\right]^{-\nu}$ to the r.h.s. of last
equation, we may write
$$\ln Z^f(\beta)=2\Gamma(\nu)\left[
\left({\beta\over 2\pi}\right)^{2\nu}E_1^{{\mu^2\over4}}(\nu,1)
-\left({\beta\over \pi}\right)^{2\nu}E_1^{\mu^2}(\nu,1)
\right],\eqno(13)$$
where now $\mu={\beta\omega\over \pi}$. Following exactly the same steps as
before, that is, using the analytical
continuation of the one-dimensional inhomogeneous Epstein function given by
(7) and taking the limit $\nu\rightarrow 0$, we get

$$\ln Z^f(\beta)= \beta\omega
+4\sqrt{{\omega\beta\over 2\pi}}\sum_{n=1}^\infty
\sqrt{{1\over n}}K_{-{1\over2}}(n\omega\beta)-4\sqrt{{\omega\beta\over
\pi}}\sum_{n=1}^\infty\sqrt{{1\over n}}K_{-{1\over2}}(2n\omega\beta)
.\eqno(14)$$
Using equations (9) and (B.1) we finally obtain,

$$\ln Z^f(\beta)=\ln\left[ {{\sinh\left(\omega\beta\right)}\over
{\sinh\left({\omega\beta\over2}\right)}}
\right]^2\;\;\;\;\Longrightarrow\;\;\;
Z^f(\beta)=4\cosh^2\left({\omega\beta\over2}\right).\eqno(15)$$

This result can be checked easily in the following way: the fermionic
oscillator
that is being considered here is the second order Grassmann oscillator studied
by Finkestein and Villasante$^{(10)}$ (in fact, we are dealing here with the
particular
case of $N=2$ of their work). In this case, it can be shown that there are
only three energies: $0$ (double degenerated),\break
$\;+ \beta\omega, \;-\beta\omega$, so that, if we
use directly the definition for $Z^f(\beta)$, we will obtain
$$\eqalignno{Z^f(\beta)&=Tr\, e^{-\beta{\hat H}^f}\cr
&=2+e^{\beta\omega}+e^{-\beta\omega}\cr
&=4\cosh^2 \left({\beta\omega\over2}\right).&(16)\cr}$$

In this paper we applied the Schwinger's formula for the one-loop
effective action to the computation of the partition function for both the
bosonic and Grassmann harmonic oscillator. Regularization by
analytical continuation was adopted. We note that depending on the boundary
condition choice ( eg. Dirichlet condition), new subtractions
(renormalizations) can be needed. But this is easily done remembering that
the Schwinger's formula contains an integration constant, which can be used to
subtract these divergent terms.

One of us (CF) would like to thank M. Asorey, A. J. Segu\'\i-Santonja
and M. V. Cougo Pinto for helpful discussions on this subject. This
work was partially supported by CAPES and CNPq (Brazilean councils of
research).

\vskip 1.0 cm

\centerline{\bf Appendix A}

Let
$$e^{\Gamma(\omega)} = \det{\hat L}_\omega\vert_{\cal F}
=\exp Tr \ln {\hat L}_\omega\vert_{\cal F},\eqno(A.1)$$
where ${\hat L}_\omega = \omega^2 - \partial_\tau^2$, and the
subscript ${\cal F}$ means that some boundary condition is assumed. Hence
$$\Gamma(\omega)= Tr\ln(\omega^2 - \partial_\tau^2)\vert_{\cal F}\eqno(A.2)$$

Taking the variation with respect to $\omega^2$ (the subscript
${\cal F}$ will be omitted but the boundary condition is understood),
we obtain
$$\eqalign{\delta_{\omega^2}\Gamma(\omega)&=Tr{\hat L}_\omega^{-1}
\delta{\hat L}_\omega\cr
&=i Tr\;\int_0^\infty ds e^{-is({\hat L}_\omega -i\epsilon)}\;
\delta{\hat L}_\omega\cr
&=\delta\biggl[-Tr\int_0^\infty\;{ds\over s}e^{-i({\hat L}_\omega
-i\epsilon)}\biggr]}\eqno(A.3)$$
Apart from an additive constant to be fixed by normalization,
integration leads to
$$\Gamma(\omega)=-Tr\;\int_0^\infty {ds\over s} e^{-is{\hat L}_\omega},
\eqno(A.4)$$
where the
$i\epsilon$ factor of convergence is tacitly assumed. Last equation
is clearly ill-defined. One way to circunvent this problem was pointed
out by Schwinger: one introduces a
cut-off $s_0$, so that the integral can be performed. Then, subtracting the
divergent terms for $s_0=0$, the remaining integral is finite.
However, there is another possible
choice$^{(5)}$, which consists in replacing expression (A.4) by
$$\Gamma(\omega)=-Tr\;\int_0^\infty ds s^{\nu-1} e^{-is{\hat
L}_{\omega}},\eqno(A.5)$$
with $\nu$ big enough, such that integral (A.5) is well defined.
Then, after the integral is made, we
perform an analytical continuation to the whole complex plane.
After subtracting the poles at $\nu=0$ (when they exist),
we take the limit $\nu\rightarrow 0$, and
this way we get a finite prescription for the integral (A.4).
In the main text we adopt this approach instead Schwinger's one.

\vskip 1.0 cm

\centerline{\bf Appendix B}

In this Appendix we shall prove that
$$\sum_{n=1}^\infty {1\over n}e^{-n\alpha}={\alpha\over2}
-\ln\left[ 2\sinh\left({\alpha\over2}\right)\right]
\;\;\; ;\;\;\; \alpha>0.\eqno(B.1)$$
With this purpose, we will define $S(\alpha)$ such that
$$S(\alpha):=\sum_{n=1}^\infty
{1\over n}e^{-n\alpha}.\eqno(B.2)$$
Diferentiating both sides of (B.2) with respect to $\alpha$ and using
the well known result of the sum of the infinite terms of a
geometrical series, we get
$$-{dS(\alpha)\over d\alpha}={e^{-{\alpha\over2}}\over
 2\sinh\left({\alpha\over 2}\right)}.\eqno(B.3)$$
Now, integrating in $\alpha$ we readly get
$$S(\alpha)={\alpha\over2}-\ln\left[ 2\sinh\left({\alpha\over2}
\right)\right], \eqno(B.4)$$
where we used that $S(\infty)=0$. This completes the desired proof.

\vskip 1.0 cm

\centerline{\bf References}
\bigskip
\item{1.} R.P. Feynman and A.R. Hibbs, {\bf Quantum Mechanics and Path
Integrals}, Mc. Graw Hill, Inc., New York, 1965.
\item{2.} For the problem at hand, this method can be found in
the following reference: H. Boschi, C. Farina and A. de Souza Dutra;
submitted for publication
\item{3.} J. Schwinger, Phys. Rev. {\bf 82} (1951) 664.
\item{4.} J. Schwinger, Lett. Math. Phys. {\bf 24} (1992) 59.
\item{5.} M. V. Cougo-Pinto, C. Farina and A. J. Segu\'\i-Santonja, to appear
in the Lett. Math. Phys..
\item{6.} M. V. Cougo-Pinto, C. Farina and A. J. Segu\'\i-Santonja,
to appear in Lett. Math. Phys..
\item{7.} G.W. Gibbons, Phys. Lett. {\bf A60} (1977) 385.
\item{8.} J. Ambjorn and S. Wolfran, Ann. Phys. {\bf 147} (1983) 1;
Elizalde and Romeo, J. Math. Phys. {\bf 30} (1989) 1133.
\item{9.} I.S. Gradstheyn and I.M. Ryzhik {\bf Table of Integrals,
Series and Products}, Academic Press, Inc. (NY), 1980, formula 8.469.3.
\item{10.} R. Finkelstein and M. Villasante, Phys. Rev.{\bf D33}
(1986) 1666.
\bye